\documentclass[aps,pra,twocolumn,10pt,superscriptaddress,showpacs,nofootinbib,longbibliography
]{revtex4-1}
\usepackage{hyperref,xcolor}

\newcommand{\QUOTE}[3]{\begin{itemize}\item[]\emph{``#1"} \\ \rule{0pt}{0pt} \hfill (#2; #3) \end{itemize}}
\newcommand{\Quote}[1]{\begin{itemize}\item[]#1\end{itemize}}

\newcommand{\demtable}{
\begin{table}[t]\caption{\label{tab:demographics}Demographic breakdowns of two populations at Bethel: students who completed Optics or Lasers (about 20 per semester) and those who earned a physics or applied physics bachelor's degree (about 20 per year) between 2010 and 2016. Data were provided by the Office of Institutional Data and Research.}
\begin{ruledtabular}
\begin{tabular}{lrr}
Gender, race, and ethnicity		& Optics/Lasers (\%) & Degrees (\%) \\ \hline
Men							& 87 & 88 \\
Women						& 13 & 12 \\
White						& 94 & 87 \\
Asian American					& 2 & 1 \\
Multiple races or ethnicities		& 4 & 6 \\
Unknown race or ethnicity			& 0 & 7
\end{tabular}
\end{ruledtabular}
\end{table}
}

\newcommand{\interviewtab}{
\begin{table*}[t]\caption{\label{tab:interviews}Pre- and post-project interviews questions, in the order they were asked during each interview.}
\begin{ruledtabular}
\begin{tabular}{p{0.005\textwidth}p{0.95\textwidth}}
\multicolumn{2}{l}{Pre-project interview questions} \\
1. & What is your anticipated major? \\
2. & What physics classes are you taking this semester? \\
3. & Have you taken any other physics lab classes besides Optics/Lasers? If so, please tell me about them. \\
4. & Have you done any undergraduate research? If so, please tell me about it. \\
5. & What do you plan to do after you graduate? \\
6. & Tell me about your final project for Optics/Lasers. \\
7. & Can you give me a little background about why this project is interesting to you? \\
8. & Why might the project be interesting to other physicists? \\
9. & In what ways do you think the project portion of the class will be different from what you've been doing so far? \\
10. & Is there anything about the project that you are excited about? \\
11. & Is there anything about the project that you are worried about? \\
12. & What do you think will be the biggest challenge when working on this project? \\
13. & Have you done any projects like this in other lab classes? \\ \\
\multicolumn{2}{l}{Post-project interview questions} \\
1. & Did you prefer the lab exercises or the final projects? Why? \\
2. & Looking back on your experience in this class, has your interest in this project changed? If so, how? \\
3. & Have you gained any insight into why this project may be interesting to other physicists? If so, tell me about it. \\
4. & What was the initial goal of the project? \\
5. & Did the goal change during the project? If so, how did it change? \\
6. & What were some especially memorable moments during this lab experience? \\
7. & What were some of the challenges, and how did they get resolved? \\
8. & What was the major outcome of your project? \\
9. & In what ways was your lab project similar to, or different from, experimental physics research? \\
10. & In what ways did your lab project prepare you to do experimental physics research? \\
11. & What are some likely next steps for continuing the project? \\
12. & Is there anything else you want to tell me about your lab project?
\end{tabular}
\end{ruledtabular}
\end{table*}
}

\newcommand{\surveytab}{
\begin{table*}[t]\caption{\label{tab:surveys}Survey questions for surveys 1, 2, and 3. Asterisks indicate Likert-style questions.}
\begin{ruledtabular}
\begin{tabular}{p{0.005\textwidth}p{0.95\textwidth}}
\multicolumn{2}{p{0.955\textwidth}}{Survey 1} \\
1. & Describe one of your project goals for the week. \\
2.\textsuperscript{*} & Who was responsible for deciding to pursue this goal? Indicate the level of responsibility for the following people: you, your lab partners, your professor. \\
3. & Please elaborate on how the decision to pursue this goal was made. \\
4. & Tell me about one specific way you anticipate contributing to this goal. \\
5. & What will you need from your lab partners in order to achieve this goal? \\
6. & What will you need from your professor in order to achieve this goal? \\ \\
\multicolumn{2}{p{0.955\textwidth}}{Survey 2} \\
1. & Describe one technical problem you encountered while working on your project
last week. \\
2.\textsuperscript{*} & Indicate the level to which you agree with the following statements: When we first encountered this problem, I felt like my lab partners and I could solve it [on our own / with help from our professor].\\
3. & Please elaborate on how you felt when you and your lab partners first encountered the problem. \\
4. & Tell me about one strategy you and your lab partners used to try to solve this problem. \\
5. & What role did your professor play in helping you try to solve this problem? \\
6.\textsuperscript{*} & Indicate the level to which you agree with the following statement: The process of trying to solve the problem was a team effort. \\
7. & Please elaborate on whether or not the process of trying to solve the problem was a team effort. \\ \\
\multicolumn{2}{p{0.955\textwidth}}{Survey 3} \\
1. & Describe a successful moment when you got something to work properly on
your project. \\
2. & In what specific ways were you able to contribute to this successful moment?\\
3. & Tell me about one of your strengths that is relevant to this successful moment or to the project more generally.\\
4.\textsuperscript{*} & Compare how you feel to now to how you were feeling before the successful moment. How have your perceptions changed as a result of the successful moment? Indicate the level of change below: Your current level of [enjoyment when working on the project / confidence in your ability to solve problems], compared to before the successful moment.\\
5. & Please elaborate on how your level of enjoyment when working on the project changed as a result of this successful moment. \\
6. &  Please elaborate on how your level of confidence in your ability to solve problems changed as a result of this successful moment.
\end{tabular}
\end{ruledtabular}
\end{table*}
}

\begin{document}


\title{Student ownership of projects in an upper-division optics laboratory course: \\ A multiple case study of successful experiences}

\author{Dimitri R. Dounas-Frazer}
\email{dimitri.dounasfrazer@colorado.edu}
\affiliation{Department of Physics, University of Colorado Boulder, Boulder, CO 80309, USA}

\author{Jacob T. Stanley}
\affiliation{Department of Physics, University of Colorado Boulder, Boulder, CO 80309, USA}


\author{H. J. Lewandowski}
\affiliation{Department of Physics, University of Colorado Boulder, Boulder, CO 80309, USA}
\affiliation{JILA, National Institute of Standards and Technology and University of Colorado Boulder, Boulder, CO 80309, USA}

\date{\today}

\begin{abstract}
We investigate students' sense of ownership of multiweek final projects in an upper-division optics lab course. Using a multiple case study approach, we describe three student projects in detail. Within-case analyses focused on identifying key issues in each project, and constructing chronological descriptions of those events. Cross-case analysis focused on identifying emergent themes with respect to five dimensions of project ownership: student agency, instructor mentorship, peer collaboration, interest and value, and affective responses. Our within- and cross-case analyses yielded three major findings. First, coupling division of labor with collective brainstorming can help balance student agency, instructor mentorship, and peer collaboration. Second, students' interest in the project and perceptions of its value can increase over time; initial student interest in the project topic is not a necessary condition for student ownership of the project. Third, student ownership is characterized by a wide range of emotions that fluctuate as students alternate between extended periods of struggle and moments of success while working on their projects. These findings not only extend the literature on student ownership into a new educational domain---namely, upper-division physics labs---they also have concrete implications for the design of experimental physics projects in courses for which student ownership is a desired learning outcome. We describe the course and projects in sufficient detail that others can adapt our results to their particular contexts.
\end{abstract}

\maketitle


\section{Introduction}\label{sec:intro}

The study and improvement of undergraduate lab courses is an increasingly important area of focus in physics education. In particular, there is an emerging emphasis on providing students with opportunities to participate in course-based projects that involve designing and conducting physics experiments. For example, the American Association of Physics Teachers (AAPT) recently endorsed a report identifying several learning outcomes for lab courses, including student competence with experimental design~\cite{AAPT2014}. More recently, an article in \emph{Physics Today} suggests that project-based approaches to lab instruction are gaining popularity in many physics departments~\cite{Feder2017b}. Similarly, the Joint Task Force on Undergraduate Physics Programs (JTUPP)  recommended that advanced lab courses incorporate multiweek research projects in order to support students' development of career-relevant technical skills~\cite{JTUPP2016}. However, compared to other physics learning environments, there is a relative dearth of education research on undergraduate lab courses~\cite{DBER2012}, making it hard to know how to productively engage students in multiweek course-based projects. In this work, we aim to provide insight into one aspect of such activities: student ownership of their projects.

What is student ownership, and why focus on this aspect of lab education? {Colloquially, student ownership of a project refers to students' feelings that the project belongs to them, and that the project outcome reflects their authentic contributions. In the education research literature, student ownership is typically mapped onto a hybrid of multiple constructs, such as students' level of autonomy, choice, control, interest, investment, or responsibility with respect to the purpose, design, implementation, or assessment of an educational activity. Some researchers are motivated to study student ownership for principled reasons. For example, in the context of educational web-based physics simulations, Podolefsky~\cite{Podolefsky2013} argued that student ownership is connected to the role of education as a tool for ``individual and collective social empowerment"\ (p.~277).} Others value ownership because they view it as a necessary ingredient for students' motivation~\cite{Milner-Bolotin2001,Enghag2006}, pride~\cite{Little2015}, or intent to persist in the sciences~\cite{Hanauer2016}.

Our interest in student ownership is also informed by numerous conversations with physics lab instructors at conferences and elsewhere. Based on these interactions, it is apparent to us that many instructors view student ownership as an important consideration for final projects in lab courses. Accordingly, we want to study the design and implementation of projects that support students in feeling meaningful levels of interest, control, investment, and responsibility. {Because we are unaware of prior work on student ownership in upper-division physics labs, there is a need for exploratory qualitative studies in this educational domain.} As a step toward this end, we performed a multiple case study of student groups who completed seven-week-long final projects in an upper-division optics lab for which student ownership is an explicit learning goal. 

In this article, we report results from both within-case and cross-case analyses of three group projects.  One of our goals is to describe the course, projects, and student experiences in sufficient detail that other researchers and instructors can determine whether and how our findings transfer to their particular contexts~\cite{Eisenhart2009}. In addition, drawing on evidence from our study, as well as from previous literature on student ownership, we make three claims: (i) coupling division of labor with collective brainstorming can help balance student agency, instructor mentorship, and peer collaboration; (ii) initial student interest in the project topic is not always a necessary condition for student ownership of the project; and (iii) student ownership is characterized by a wide range of emotions that fluctuate in time as students alternate between extended periods of struggle and moments of success while working on their project.

This paper is organized as follows. In Sec.~\ref{sec:background}, we summarize relevant background literature on optics education, physics projects, and student ownership. In Sec.~\ref{sec:framework}, we define what we mean by ``student ownership of projects." In Sec.~\ref{sec:context}, we describe the institutional, departmental, and course context for our study. Next, in Sec.~\ref{sec:methods} and Sec.~\ref{sec:results}, we describe our case study methodologies and present our results. {In Sec.~\ref{sec:discussion} we discuss the limitations and implications of our work. Finally, we summarize our findings and suggest areas for future work in Sec.~\ref{sec:conclusion}.}


\section{Background}\label{sec:background}

Three areas of education research are relevant to our study: teaching and learning in optics courses, multiweek projects in lab courses, and student ownership in science courses. We summarize relevant work from each area, with an emphasis on literature related to undergraduate physics education.

\subsection{Teaching and learning in optics courses}

Previous research on optics education spans a variety of topics, including the development of multimedia activities~\cite{Zhang1997}, online materials~\cite{Mateyick2005}, and interactive learning strategies meant to be implemented in lecture courses~\cite{Jones2015} or hybrid lecture-studio courses~\cite{Sorensen2011}. {In a course for preservice teachers, Atkins and Salter~\cite{Atkins2010} described students' processes for constructing definitions of ``blurriness."} Other work has focused on characterizing students' conceptual difficulties~\cite{Colin2001} and problem-solving strategies~\cite{Undreiu2008} in theory-based geometrical optics contexts, as well as their use of model-based reasoning when completing an experimental optics task~\cite{Zwickl2015b}. In the context of upper-division optics labs, many sequences of activities have been described in detail. Activity topics include laser physics~\cite{Henningsen2011}, interference of correlated photons~\cite{Galvez2005}, single-photon experiments~\cite{Pearson2010}, {ultrafast optics~\cite{Hoyt-unpublished},} and spectroscopy~\cite{Blue2010}. Additionally, multiple course transformation efforts have been documented~\cite{Blue2010,Zwickl2013,Masters2010}. Some of these efforts incorporated final projects~\cite{Zwickl2013,Masters2010}; however, the project portions of the transformed courses have been neither described nor studied in detail.

Whether developing curricular materials, characterizing student reasoning, or transforming courses, most prior work has focused on particular optics concepts or skills. The course transformation documented by Masters and Grove~\cite{Masters2010} is an exception; the purpose of that effort was to ``combine the goals of developing conceptual understanding and laboratory independence."\ (p.~486). However, while students' ability and desire to work independently on lab activities is related to their sense of ownership of those activities, independence and ownership are distinct constructs (see Sec.~\ref{sec:framework}).

Two recent interview studies have focused on understanding optics education through the lenses of lab instructors~\cite{Dounas-Frazer2017arXiv} and members of the photonics workforce~\cite{Zwickl2015a}. We recently investigated lab instructors' perceptions about whether and how students engage in model-based reasoning during optics lab activities~\cite{Dounas-Frazer2017arXiv}. Many instructors in that study said that iteration was an important aspect of experimentation, and almost all described multiple ways that students iteratively improve their experiments (e.g., making changes to the apparatus or data-taking procedures). These findings suggest that iteration is a common and important feature of many optics lab activities. Zwickl et al.~\cite{Zwickl2015a} explored the perceptions of employees in the photonics workforce about the skills required to succeed in the workplace. Participants in that study indicated that computation and communication (i.e., written documentation and oral presentations) are important professional skills. Participants further suggested that these and other relevant skills are commonly learned during academic coursework, undergraduate research experiences, internships, and on-the-job training. Although neither of these studies~\cite{Dounas-Frazer2017arXiv,Zwickl2015a} speaks directly about multiweek projects in formal lab courses, they imply that research-style projects that incorporate iteration, documentation, and oral presentations may be particularly beneficial for students interested in pursuing optics research or career pathways.

\subsection{Multiweek projects in physics lab courses}

Several studies have documented the benefits of multiweek projects in undergraduate physics labs. Holmes and Wieman~\cite{Holmes2016} found that students who completed a design-based lab course described engaging in many of the tasks associated with table-top experimental physics research. Similarly, Juma et al.~\cite{Juma2010} found that, after completing a capstone project in an advanced electronics lab course, students' self-reported learning outcomes included  improved competence with general experimentation tasks (e.g., troubleshooting), as well as a better understanding of the concepts and equipment related to their project. And, in a study of a final project in an introductory physics course for non-physics majors, Martinuk et al.~\cite{Martinuk2009} argued that projects may have improved students' confidence with certain estimation computations, but that students' tendency to apply physics principles to ``everyday problems" did not improve.

Other work has focused on noncognitive impacts of projects. In a study of students enrolled in an advanced lab course, Irving and Sayre~\cite{Irving2014} argued that student collaboration on ``long and difficult physics experiments" facilitated students' introduction to the ``authentic expectations, practices, content knowledge, and discourses" of practicing physicists\ (p.~14). Quan and Elby~\cite{Quan2016} explored non-cognitive aspects of semester-long research projects. They showed that some students experienced coupled shifts in both their views about the nature of science and their beliefs about their ability to contribute to research. In addition, in the context of open-ended multiweek projects in a lab course focused on contemporary experimental physics, Eblen-Zayas~\cite{Eblen-Zayas2016} showed that metacognitive activities and in-class discussions that were ``intended to normalize the feelings that students had when progress was slow or frustration grew" had positive impacts on measures of students' enjoyment and confidence with respect to experimentation. Together, these studies~\cite{Holmes2016,Juma2010,Martinuk2009,Irving2014,Quan2016,Eblen-Zayas2016} suggest that multiweek projects can support students' development as competent and confident physicists.

Some practitioner-oriented work also exists. For example, Planin\v{s}i\v{c}~\cite{Planinsic2007} and Gandhi et al.~\cite{Gandhi2016} have described introductory physics courses that include multiweek final projects. In particular, the course described by Gandhi et al.\ was informed by Papert's~\cite{Papert1991} constructionist model of learning---i.e., ``learning by making"---and was designed in an educational context that values iteration, collaboration, and student ownership~\cite{Albanna2013}. However, Gandhi et al.\ did not define, operationalize, or measure student ownership. {While we are unaware of work that focuses on student ownership of projects in physics lab courses, there is nevertheless a growing body of literature on ownership in physics and other science courses.}

\subsection{Student ownership in science courses}\label{sec:ownership}

To our knowledge, the first empirical studies of student ownership in physics contexts were described in the dissertations of Milner-Bolotin~\cite{Milner-Bolotin2001} and Enghag~\cite{Enghag2006}. Milner-Bolotin's work focused on nonscience majors working on semester-long projects; she found that student ownership and motivation were interrelated phenomena~\cite{Milner-Bolotin2001}. Enghag and colleagues~\cite{Enghag2006,Enghag2008,Enghag2009} focused on future teachers working on two-week-long miniprojects~\cite{Enghag2008} and aeronautical engineering students working on context-rich ``short story" problems~\cite{Enghag2009}. In these studies, Enghag and colleagues developed models of student ownership at the individual and group levels. More recently, some work has focused on ownership and autonomy in large-enrollment introductory physics courses. For example, Demaree and Li~\cite{Demaree2009} described pedagogical approaches in a course for which a major goal was ``to have students take ownership for their knowledge development."\ (p.~125). Hall and Web~\cite{Hall2014} found that higher levels of student autonomy---an important aspect of student ownership---were correlated with better motivational and affective experiences in the course, as well as higher course grades.

Beyond the domain of physics education, a sequence of studies in biology are particularly relevant for our work. Hanauer et al.~\cite{Hanauer2012} performed a linguistic analysis of interviews with biology students in a traditional lab course, inquiry-based lab course, and independent research course. They identified several features of the learning environment that facilitate student ownership of projects, including personal agency and appropriate mentorship. In a follow-up study, Hanauer and Dolan~\cite{Hanauer2014} developed and evaluated the Project Ownership Survey (POS), a Likert-style assessment of student ownership. The POS distinguished between the levels of ownership experienced by students in a traditional biology lab versus those in a research lab, with higher levels in the latter  case. Most recently, Hanauer et al.~\cite{Hanauer2016} have performed preliminary validation of a survey designed to measure student persistence in the sciences; early results suggest that POS scores, among other variables, predict biology students' intent to become research scientists. {The work of Hanauer and colleagues~\cite{Hanauer2012,Hanauer2014} informed our operationalization of student ownership in the present study.}

Finally, our study was also informed by work from beyond the domain of science education. Wiley~\cite{Wiley2009} performed a literary analysis of discussions of student ownership in the education literature. Wiley found that many practitioners neglect to define the concept of student ownership, sometimes shifting between various implicit definitions:
\Quote{
There is a large supply of practitioner articles in which authors use ownership without definition or explication. Most of these articles focus on a specific program or activity and mention \emph{ownership} only in passing. \ldots\ Likewise, the term \emph{ownership} is often applied in a manner that seems to slip and slide among various meanings. By not bothering to define it strictly, authors sometimes unconsciously shift among meanings even within a single article.\ (italics in original; pp.~7--9)
}
To avoid these pitfalls in the present work, we define and operationalize our conception of student ownership in the following sections.


\section{Student ownership of projects}\label{sec:framework}

Our conception of student ownership is heavily informed by the work of Wiley~\cite{Wiley2009}, Hanauer and colleagues~\cite{Hanauer2012,Hanauer2014}, Milner-Bolotin~\cite{Milner-Bolotin2001}, and Enghag and colleagues~\cite{Enghag2006,Enghag2008,Enghag2009}. In this section, we draw on this and other work to both define student ownership and describe how it evolves over time.

\subsection{Right/responsibility, buy-in, and identification}

Based on his literary review, Wiley~\cite{Wiley2009} articulated a three-part definition of ownership. Student ownership may refer to students' freedom to make decisions about the nature of their education, and their responsibility for the outcomes of their efforts. Alternatively, student ownership may refer to students' buy-in, i.e., their commitment to, and investment in, an activity. Similar to owning stock in a company, a student can be invested in an activity even if they do not have control over it. Finally, student ownership may refer to students' identification with an activity. In this case, ownership may manifest as a sense of ``pride over, intense commitment to, or a personal connection with" the activity\ (p.~19). Thus, according to Wiley, \emph{student ownership refers to students' right and responsibility, buy-in, or identification with an activity or environment}.

\subsection{Interaction between student and environment}\label{sec:dimensions}

Hanauer and colleagues~\cite{Hanauer2012,Hanauer2014} expanded Wiley's definition by identifying several features of projects that both support student ownership and indicate its presence. In doing so, they argued that student ownership must be understood as resulting from a ``complex interaction between the student and the educational environment."\ \cite{Hanauer2012} (p.~379).  Their work suggests that students' sense of project ownership has five dimensions, upon which we elaborate below.

\emph{First, students may feel ownership when they have personal agency in the project.} In particular, Hanauer et al.~\cite{Hanauer2012} stressed the importance of student input on research questions and strategic decisions about project execution. Many researchers have described the connection between student ownership and students' level of choice, control, and responsibility with respect to a project~\cite{Dudley-Marling1995,Podolefsky2013,Milner-Bolotin2001,Enghag2006,Enghag2008,Enghag2009,Savery1996,Rainer2002}; hence this dimension of ownership is related to Wiley's definition of ownership as right and responsibility. {Importantly, personal agency does not imply unconstrained choice or unregulated autonomy.} For example, Milner-Bolotin~\cite{Milner-Bolotin2001} found that, compared to students who chose their project topics, students whose project topic was assigned by an instructor were neither more nor less likely to feel ownership of their projects at the end of the semester. She further argued that student ownership ``may be also more related to [students'] opportunity to choose their group members than topic choice," indicating that students' control over the team is an important aspect of student agency\ (p.~103). Additionally, Dudley-Marling~\cite{Dudley-Marling1995} cautioned that ``[w]orking independently---with limited teacher support and direction---is a perverse notion of ownership."\ (p.~11). Instead, Dudley-Marling argued that student ownership requires striking a balance between teacher support and student control. The importance of such a balance is also reflected in the work of Hanauer and colleagues.

\emph{Second, students may feel ownership when they are able to solicit assistance and direction from a mentor.} According to Hanauer and colleagues~\cite{Hanauer2012,Hanauer2014}, mentorship must be combined with student agency, being neither overly prescriptive nor insufficiently supportive. Similarly, Savery~\cite{Savery1996} argued that ``hands-off" teaching does not promote ownership. Rather, teachers should meet with student groups to address problems as they arise, making sure that the responsibility to decide upon and enact solutions resides with the students. Mikalayeva~\cite{Mikalayeva2016} framed the ideal student-instructor relationship as a form of ``cooperative dominance": student ownership arises when instructors coach students and provide authoritative-but-restrained guidance, not when they relinquish control and become passive participants in the learning process. A slightly different vision for instruction was offered by Demaree and Li~\cite{Demaree2009}. They described the role of the instructor as a broker rather than an authority figure, i.e., as ``an agent that acts between two communities, and in our case, attempts to help guide the classroom community closer to that of the practicing physics community"\ (p.~126). Thus, Wiley's definition of ownership as right and responsibility involves constraints that arise from balancing student autonomy and instructor guidance.

\emph{Third, students may feel ownership when they collaborate with peers to overcome challenges.} Hanauer and colleagues~\cite{Hanauer2012,Hanauer2014} noted that social interaction with peers can complement mentorship by an instructor when students work to overcome problems on their project. Savery~\cite{Savery1996} and Milner-Bolotin~\cite{Milner-Bolotin2001} also argued that teamwork is an important social component of ownership. Moreover, in their framework for encouraging ownership in teacher education, Rainer and Matthews~\cite{Rainer2002} emphasize the importance of blending both independent and collaborative investigation: teams can function effectively by identifying and leveraging the expertise of individuals in the group, making decisions together about how to organize their time, and sharing their ideas with one another. Importantly, Enhag and Niederrer~\cite{Enghag2008} differentiated between group ownership and individual ownership. Group ownership may occur ``when the students together with the teacher decide on the management of the task," including decisions about the makeup of the group, how various tasks will be executed (and by whom), and how the results of the project will be presented (p.~634). Individual ownership, on the other hand, may occur when a student contributes a particular idea that is taken up by the group as a whole~\cite{Enghag2008}. Therefore, in addition to balancing student autonomy and instructor guidance, the conception of ownership as right and responsibility also involves negotiation of control over various aspects of the project among group members.

\emph{Fourth, students may feel ownership when they are interested in the project and perceive it as valuable to others.} Hanauer and colleagues argued that projects that connect to students' personal history, major social problems, or issues that are relevant to the broader scientific community may be particularly significant to students.  Indeed, Milner-Bolotin~\cite{Milner-Bolotin2001} found that students' initial interest in a project resulted in student ownership early on in the project. Along similar lines, some teachers encourage student ownership by engaging their students in work that they (the students) find ``purposeful"~\cite{Rainer2002}. This dimension of project ownership is related to Wiley's definition of ownership as buy-in; students are likely to invest their own resources---including personal time~\cite{O'Neill2005}---in projects that are interesting, valuable, or relevant to science or society.

\emph{Fifth and last, students may feel ownership when they feel excited about the scientific process, willing and able to contend with problems as they arise, and satisfied with their achievements.} According to Hanauer and Dolan~\cite{Hanauer2014}, student ownership is facilitated by strong emotional connections between the student and their project, including ``genuine excitement for the process of scientific inquiry" and moments of ``pride, happiness, or satisfaction" upon achieving a specific finding or discovery\ (pp.~150--151). Others have also drawn connections between ownership and positive emotional responses to a project. For example, Little~\cite{Little2015} argued that student ownership is an important component of feeling proud of physics projects in both research and educational settings\ (p.~9), and O'Neill and Barton~\cite{O'Neill2005} argued that urban middle school students express ownership in science when they express positive views about themselves. Moreover, Hanauer et al.~\cite{Hanauer2012} found that students' ``ability and willingness to contend with problems that arose within the scientific inquiry process" was also a hallmark of student ownership\ (p.~384). Connections between students' ownership of a project and their intrinsic motivation have been documented by other researchers as well~\cite{Savery1996,Milner-Bolotin2001,Enghag2006,Enghag2008,Enghag2009}. This dimension of ownership is related to Wiley's definition of ownership as identifying with the learning experience, which involves emotional connection or intense commitment to an activity.

Together, the work of Wiley~\cite{Wiley2009} and Hanauer and colleagues~\cite{Hanauer2012,Hanauer2014} forms our understanding of what student ownership is and how it relates to various features of the learning environment. However, student ownership can develop and fluctuate over time~\cite{Milner-Bolotin2001,Enghag2006,Enghag2008,Enghag2009}. Therefore, a complete understanding of student ownership must also take into account its evolution in time.

\subsection{Dynamic processes}

As an example of the dynamic nature of student ownership, we focus on Milner-Bolotin's study of non-physics majors working on semester-long projects in an introductory physics course~\cite{Milner-Bolotin2001}. Milner-Bolotin described several temporal patterns in students' sense of ownership. She found that, while the ability to choose a project topic led to high ownership early in the semester, the impact of topic choice declined as the semester progressed. Initially high levels of student ownership dropped during the middle of the semester and increased again toward the end of the semester, regardless of whether students were able to choose their project topic.

Milner-Bolotin attributed these dynamics to the challenges of teamwork, division of labor, and time management. Students' initial enthusiasm for the project was replaced with frustration in the face of the realities of project execution. However, as students began to see ``the fruits of their hard work," their interest and investment in the projects increased\ (p.~140). In particular, Milner-Bolotin~\cite{Milner-Bolotin2001} noted that final presentations were particularly impactful for students. Many students reported feeling surprised and encouraged by their progress, and they wanted to share their work with their classmates and friends. These findings suggests that interest, collaboration, and emotional connections to the project are interrelated phenomena that change over time. Hence, \emph{students' development of a sense of project ownership evolves in complex and non-monotonic ways}.

In summary, student ownership refers to students' responsibility for, investment in, or identification with a project~\cite{Wiley2009}. Students may feel ownership when they (i) have personal agency in the project, (ii) have access to appropriate mentorship, (iii) collaborate with peers on challenging problems, (iv) perceive the project to be interesting or valuable, and (v) feel excited about the process, capable to solve problems, and satisfied with their achievements~\cite{Hanauer2012,Hanauer2014}. Last, students' sense of project ownership fluctuates over time, decreasing during times of challenge and increasing when their hard work results in moments of success~\cite{Milner-Bolotin2001}.  In this study, we explore whether and how students in an upper-division optics lab developed a sense of ownership of their final projects.


\section{Context}\label{sec:context}

Our study was performed at Bethel University (hereafter, ``Bethel"), a medium-sized, more selective, predominantly white, private not-for-profit Christian college~\cite{Carnegie2015}. The physics department at Bethel ranks among the ten largest undergraduate-only programs in the country~\cite{AIP2015}. On average, 20 people earn a physics bachelor's degree from Bethel each year. A breakdown of the gender, race, and nationality of degree recipients is provided in Table~\ref{tab:demographics}.

\subsection{Course context}

Our study focuses on the final project portion of two similar physics courses taught at Bethel: Optics and Lasers. Both courses are required for some physics bachelor's degree tracks at Bethel. {Typical enrollment in each course is about 20 students, most of whom are physics or engineering majors}; a demographic breakdown of Optics and Lasers students is provided in Table~\ref{tab:demographics}.\footnote{We report demographic data for two reasons: to enable metastudies of research contexts and participant populations in the physics education literature, and to facilitate potential future comparisons between our work and other studies of similar phenomena in different populations.} The courses are offered in alternating spring semesters. {Students enrolled in Optics and Lasers are typically in their third or fourth year of coursework.} Topics covered in Optics include waves, electrodynamics, light propagation, geometrical optics, superposition, interferometry, polarization, diffraction, and an introduction to lasers. Topics covered in Lasers include laser light properties, laser output modification, and various types of laser systems: external cavity, diode, dye, gas, semiconductor, fiber, and solid state. In both courses, experiments are performed on large optical tables, students regularly engage in computer-aided data analysis, and they have access to a variety of modern instruments and equipment: oscilloscopes, power meters, spectrum analyzers, synthesizers, and multiple types of laser systems.

{While the topics covered in Optics and Lasers are different, both courses are taught by the same instructor and the overall goals and structure of the courses are similar---including the format and topics of final projects}. In both courses, the syllabus emphasizes student autonomy as one of the course outcomes: ``We will relentlessly work toward the scenario in which you formulate your own ideas on paper and in the lab." In particular, the syllabus for Optics frames the course as an authentic experimental physics experience in which unexpected challenges and troubleshooting are to be expected:
\Quote{
This class also aims to help prepare our physics, applied physics, and engineering majors for research. The laboratory often involves considerable freedom to try varying approaches with very challenging goals. Things will not ``work right" when you start, and you may be ``thrown into" research level areas for which you are initially ill prepared. That is the nature of real research and development.
}
Because the Optics and Lasers courses are so similar, we refer to them collectively as ``Optics/Lasers."

When asked to articulate learning outcomes for Optics/Lasers, the instructor identified six goals. Five of these goals were for students to develop (i) a deep understanding of the relevant physics, especially with respect to connections between theory and experiment; (ii) an appreciation of the importance of lasers and optics in science and industry; (iii) the ability to think like a physicist, especially with respect to the ability to use estimation and scaling relationships in order to engage in ``back-of-the-envelope" reasoning; (iv) technical skills related to equipment operation, apparatus design, and data analysis; and (v) confidence in their experimental abilities and clarity about career choices within physics and engineering. The sixth goal---which is particularly relevant to our study---was for students to feel engagement, fire, and ownership. When elaborating on this goal, the instructor indicated that he wanted students to pursue their own ideas, spend additional time in the lab beyond what is expected, engage in animated discussions with classmates and instructors, and generally feel excited.

{Optics and Lasers are both 14-week courses that consist of two halves, each with a lecture and lab component. Lectures span both halves of the course. During the first 7 weeks, lab activities are guided by lab manuals and focus on a particular optical phenomenon. There are no guided activities during the second half of Optics/Lasers. Instead, the final 7 weeks of each course are dedicated to final projects. These projects are the focus of our study.}

\demtable

\subsection{Project context}

{During the project portion of the Optics/Lasers, a second instructor joins the lead instructor of the course to help with mentorship of student groups.} Students are presented with a portfolio of possible projects. They rank each project according to their interest, and the instructors assign groups according to students' project preferences. Groups consist of 2 to 4 students. There are no lab manuals to guide students on their projects, and projects do not take place in a dedicated classroom or during regularly scheduled times of day. Rather, the instructors provide students with relevant scientific journal articles and, when available, documentation written by students from previous semesters who worked on an earlier phase of the same experiment. Projects take place in various research labs throughout the physics department, and groups coordinate their own time management and division of labor. {Students typically spend 5 to 10 hours per week working on their projects during the first few weeks of the project; during the last few weeks, students work on their projects for about 10 to 15 hours per week.} Projects ultimately culminate in oral presentations and reports written in the style of a journal article.

Project topics are informed by the instructors' research interests and available equipment in the department. For example, the projects we describe in this paper involved the development of a frequency comb laser, a scanning spectrometer, and a surface plasmon laser. The instructors collaboratively brainstorm project topics and rely on student groups to design initial versions of apparatuses and experimental setups. Most projects are longterm, spanning many semesters; these projects are carried forward by groups of students in Optics/Lasers, and by students for whom the project is part of their senior thesis requirement or their on-campus summer research experience. When students work on a project that is a continuation from the past, they may rely on a previous group's final report or project notebook for guidance. They may also discuss the project with students who had completed the course in a previous semester. In addition, the achievements of a previous group may be used as a benchmark for improvement; groups in Optics/Lasers are sometimes challenged to make an existing apparatus more accurate, more versatile, or more portable compared to what a previous group accomplished. When students work on projects that will continue into the future, the instructors emphasize that their notebooks must be sufficiently detailed and well-organized that they (the notebooks) will be useful to future student groups. Thus, for many students in Optics/Lasers, their engagement with the final project is informed by its situation as part of a longterm research endeavor.


\section{Methods}\label{sec:methods}

Because there is a dearth of research on ownership in physics labs, we opted for an in-depth qualitative methodology. {In particular, we conducted a multiple case study of three student projects in Optics/Lasers in order to provide a detailed description of several processes that support student ownership}. We performed both within-case and cross-case analyses. Within-case analyses were descriptive. During the cross-case analysis, we used both \emph{a priori} and emergent coding schemes to look for thematic patterns across cases.

Our case study methodology was informed by the work of Stake~\cite{Stake1978}, Merriam~\cite{Merriam1998}, and Creswell~\cite{Creswell2007}. Stake argues that case studies constitute a particularized knowledge that complements the more generalized knowledge represented by models or laws. However, though case studies do not (and are not meant to) constitute generalized knowledge, the reader may nevertheless determine whether and how features of the case apply to their own experiences~\cite{Stake1978}. Merriam stresses the importance of clearly articulating the boundaries of the case. Moreover, in order to construct a comprehensive description of a given case, Merriam recommends extensive data collection that draws on multiple sources of information, such as observations, interviews, documents, and videos~\cite{Merriam1998}. Finally, Creswell notes that, while studies of multiple cases inevitably dilute the overall description of each case, they provide multiple perspectives on a single issue and allow one to identify themes across cases. In such studies, it is important to be purposeful in selecting which cases to analyze; representative cases are the most useful for abstracting across cases~\cite{Creswell2007}.

In this section, we describe our participants, case boundaries, data sources, case selection criteria, and methods for both types of analyses (i.e., within-case and cross-case).

\subsection{Participants}

We collected data from one semester each of Optics and Lasers (two semesters total). Across both semesters, 34 unique students completed 12 distinct final projects. All students agreed to participate in the study. Two students completed projects in both semesters, resulting in a total of 36 participants. {The race, ethnicity, and gender of participants closely matched the historical demographic data for the Optics/Lasers course presented in Table~\ref{tab:demographics}.} All participants were volunteers; those who completed interviews received a small monetary incentive.  We have reported on data collected from these participants in a different study~\cite{Stanley2016}. In this study, we selected 3 projects for in-depth analysis.

\subsection{Case boundaries}

\interviewtab

\surveytab

Stake~\cite{Stake1978} describes case studies as the study of bounded systems, and argues that such studies give ``great prominence to what is and what is not `the case'---the boundaries are kept in focus." (p.~7). In our multiple case study, each case is a distinct student project in the Optics/Lasers course. Temporally, each case is bounded by two distinct episodes: the case ``started" when the professor described the project options to the students, students submitted their ranked project choices to the professor, and the professor assigned students to the project; and the case ``ended" after the group gave their final presentation and submitted their final report.

The case boundaries also limit inclusion in the case to the two or three students who worked together on a particular project. However, while other people are not included in the case, students' interactions with others may fall within the case boundaries if those interactions inform students' engagement with the project. For example, mentorship-style interactions between the students and the Optics/Lasers professors are included in the case even though the professors are not. Moreover, each case focuses on the actions and interactions of the students in the context of the project. The students in our study likely interacted with one another in many other contexts, including as friends, housemates, or classmates in other courses. Such interactions are beyond the scope of the case.

Although each case is a group of students, we do not distinguish between individual and group ownership of a project, a type of distinction that has been emphasized in other studies~\cite{Enghag2009,Enghag2008,Enghag2006}. Our study was not originally designed to investigate this distinction, and students' fluid use of both singular and plural pronouns (e.g., ``I, me, my" and ``we, us, our") makes it difficult to make strong claims about such distinctions in retrospect. Hence, one limitation of our study is the conflation of individual and group ownership of a project.

\subsection{Data sources}

Consistent with the recommendations of Merriam~\cite{Merriam1998}, we collected and analyzed multiple forms of data: electronic copies of notebook entries and final reports, pre- and post-project interviews, and weekly project surveys. Data collection was facilitated by the support and cooperation of the lead instructor for Optics/Lasers. For example, the instructor compiled student artifacts, helped coordinate interviews, and offered a small amount of course credit for completion of weekly surveys.

For each of the 12 groups in Optics/Lasers, we collected the project notebook and the final report. Notebooks and reports were generated collaboratively by all group members. Notebooks ranged from 10 to 40 pages in length, and reports were typically about 10 pages long.

Pre-project interviews consisted of 13 questions that focused on students' preparation for, interest in, and expectations about the project. Post-project interviews consisted of 12 questions that focused on students' interest in and experiences with the project, as well as their perceptions of how the project was related to experimental physics research. The interviews were semi-structured, and the prompts are provided in Table~\ref{tab:interviews}. Many of our interview questions were inspired by (and some are identical to) those developed by Hanauer et al.~\cite{Hanauer2012}. Interviews were conducted by the first and last authors, either in person or remotely via videoconference. Interviewers occasionally deviated from the protocol in order to ask participants to clarify or elaborate on an idea. Out of 36 total students who completed projects in Optics/Lasers, 33 participated in interviews before the projects started, and 35 after they ended. Pre- and post-project interviews lasted 10 to 35 minutes; we collected a total of 19 hours of audio data. All interviews were transcribed, and the transcripts are the data that we analyzed.

During the last 6 weeks of the project, we administered weekly surveys to all Optics/Lasers students. We designed three different surveys; surveys 1, 2, and 3 focused, respectively, on students' goals, challenges, and successes while working on the project. Surveys were administered at the end of each week, to be completed over the weekend. No survey was administered after the first week because students were still familiarizing themselves with the nature of the project and the corresponding equipment. To avoid repetition, we cycled through the three surveys such that surveys 1, 2, and 3 were administered successively in two three-week cycles.

Each survey consisted of 5 free-response questions. In addition, each survey included one or two Likert-style questions whose purpose was to help guide students' responses to subsequent free-response items. Survey 1, which focused on students' project goals, was designed to provide insight into student agency, instructor mentorship, and peer collaboration. Survey 2, which focused on challenges, was designed to provide insight into instructor mentorship, peer collaboration, and affective responses. Last, Survey 3, which focused on successes, was designed to provide insight into student agency and affective responses. Importantly, the surveys did not probe students' interest in their project or their perceptions of its value. We deliberately omitted these dimensions of ownership in order to avoid a scenario in which the surveys repeatedly drew attention to a student's perception of the project as uninteresting or not valuable. Hence, the surveys probed several (but not all) aspects of project ownership multiple times in different contexts over the course of the project. Survey items for all three surveys are presented in Table~\ref{tab:surveys}. In total, 199 surveys were completed, corresponding to an overall survey completion rate of over 90\%.

\subsection{Case selection}

Out of the 12 groups who completed projects in Optics/Lasers, we selected 3 for analysis: group A, the frequency comb group (2 students); group B, the spectrometer group (3 students); and group C, the plasmon laser group (2 students). Selection of these groups resulted from applying the following three filters to our data set: first, eliminate groups with more than 3 students; second, eliminate groups for whom at least 1 interview or 2 surveys were not completed; and third, based on a preliminary analysis of post-project interviews, eliminate groups in which at least 1 student described having a negative experience. The first criterion ensured that we focused on the smallest groups in our dataset, which would help mitigate the dilution of detail with which any particular case can be described. The second criterion ensured that we had a complete (or nearly complete) data set for each case, facilitating in-depth case descriptions. And the third criterion increased the likelihood that all students in each case developed a sense of ownership of the project, minimizing complications in data interpretation that may arise due to conflation of individual and group ownership.

After applying these three filters to our data set, four eligible cases remained. To further reduce the number of cases, we performed a preliminary analysis of pre-project interviews. Three cases comprised students who all had similar backgrounds and preparation: both students in group A had extensive optical physics research experience; all three students in group B were interested in engineering-type projects; and both students in group C were uncertain about their plans beyond graduation, and neither had prior research experience. Students in the fourth group were more heterogeneous in terms of their backgrounds and preparation, and hence this case was discarded from further analysis. Thus, our multiple case study focused on groups A, B, and C.

When discussing purposeful sampling of cases, Creswell~\cite{Creswell2007} endorses the selection of ``ordinary cases, accessible cases, or unusual cases."~(p.~75). In our study, groups A, B, and C represent ordinary cases within the context of Optics/Lasers; indeed, in previous work, we argued that most students in our broader participant pool felt ownership of their projects~\cite{Stanley2016}. Nevertheless, the context of Optics/Lasers is itself unusual. Our experience designing, teaching, and studying lab courses leads us to believe that the course features described in Sec.~\ref{sec:context} are both atypical and exemplary---and therefore worth describing in detail.

\subsection{Within-case analyses}

Our descriptive within-case analyses followed several recommendations by Creswell~\cite{Creswell2007} and Merriam~\cite{Merriam1998}. For example, Creswell~\cite{Creswell2007} emphasizes that data collection and analysis procedures should be replicated for each case when drawing on multiple cases to illustrate a single issue, and one analysis strategy recommended by Merriam~\cite{Merriam1998} involves producing a detailed description of each case by constructing a chronology of key issues. Our data collection procedures were nearly identical for all students in Optics/Lasers, and hence for all three groups in our study. Here, we describe our process for constructing descriptive case chronologies.

For each case, data sources were grouped into three categories according to when the data were collected: the beginning, middle, and end of the project. Thus pre-project interviews were in one category (beginning), surveys and notebooks in another (middle), and post-project interviews and final reports in the third category (end). The first author began constructing a case log by reading and summarizing first the end data, followed by the beginning data, and then the middle data. This order was chosen to give the researcher insight into what the group actually accomplished, what the group's ``initial conditions" were, and how the group navigated from the initial conditions through to project completion.

After summarizing the data, two to four key issues were identified.{``Key issues" were technical issues that were discussed by multiple group members or in multiple contexts (i.e., reflections, post-project interviews, and final reports).} Identification of key issues was especially informed by questions in the post-project interviews and weekly surveys that asked students to describe specific events related to the project. Corresponding dated notebook entries allowed for the construction of a detailed case chronology, or log. This log identified each issue using a short descriptor (e.g., ``achieving a mode-lock"). Relevant summaries and excerpts from each data source were grouped under the corresponding descriptor. The case log was then presented to, and discussed with, the research team as a whole.

For each case, our within-group analysis constitutes a descriptive and chronological summary of the case log. These chronological descriptions facilitate understanding of the complexity of each case. Such understanding provides important context for interpreting the results of our thematic cross-case analysis.

\subsection{Cross-case analysis}\label{sec:cross-case-methods}

We focused only on three data sources for our cross-case analysis: pre-project interviews, weeky survey responses, and post-project interviews. In total, we analyzed 14 interviews and 39 survey responses (in each of groups A, B, and C, one student failed to complete one of the weekly surveys). To analyze these data, we used the following \emph{a priori} coding scheme, which was inspired by the five dimensions of project ownership identified by Hanauer and colleagues~\cite{Hanauer2012,Hanauer2014} (Sec.~\ref{sec:dimensions}):
\begin{enumerate}
\item \emph{Student agency.} Our operationalization of agency was informed by the work of Bandura~\cite{Bandura2008}. This code was assigned when a student described their own participation in setting goals, anticipating future events, forming plans and strategies, or reflecting on progress, contributions, or the meaning of the project. This code was also assigned when the student described management of efforts. We did not distinguish between individual and group agency, and therefore we did not limit this code to instances where the student was the only contributor to, e.g., setting goals or forming plans. Examples include: ``I am looking for other ways [to solve the problem],"  ``We discussed the pros and cons of rearranging our setup," and ``We were talking with our group members and decided [on a course of action]."
\item \emph{Instructor mentorship}. This code was assigned when a student described interacting with one of their professors in a mentorship capacity. Examples include: ``My partner and I met with the professor and got an idea of what all needs to be done for this project," ``We needed to confirm with the prof that our setup was looking good," and ``Our professor was able to find us another photo-diode."
\item \emph{Peer collaboration}. This code was assigned when a student described working on the project with other students who were included in the case boundary (i.e., their group members). Examples include: ``The whole group was contributing thought into solving the problem," ``Processes ended up distributed amongst the group members," and ``We both worked together to find the right equipment to accomplish the task."
\item \emph{Interest and value}. This code was assigned when a student described the extent to which they perceived the project to be personally interesting or of value to others, including other people in the course, the physics department, the university, or relevant scientific and educational communities. Examples include: ``I think it'll take a little bit of learning to find some interest in [the project]," ``The 3D printing stuff is really interesting to me," and ``The inexpensiveness and the nano scale size of it would be appealing to other people."
\item \emph{Affective response}. This code was assigned when a student described an emotional response to a particular issue, accomplishment, setback, or pattern of events on the project. Examples include students describing particular events as ``frustrating," ``tedious," ``disappointing," ``overwhelming," ``exciting," ``a relief," or ``a boost in morale." We also assigned this code to instances where the student described the extent to which they believed they were able to complete a task, achieve a goal, or contend with challenges. Examples include students saying they felt ``worried," ``nervous," ``unsure," ``like no progress can be made,"  ``confident," ``able to solve those problems," or like ``we always knew we would [solve the problem]."
\end{enumerate}

We used a multi-pass coding process to analyze interviews and survey responses using the following approach. First, for each of the five \emph{a priori} code categories, the first author read through all transcripts and survey responses, identifying excerpts related to the corresponding code category. Thus, each transcript and survey response was read in its entirety a total of five times. Some excerpts received multiple codes. For example, instances where students described their own participation in collaborative decision making processes were coded as both \emph{student agency} and \emph{peer collaboration}.

Next, for each category, the second author read through all the coded excerpts to verify that they matched the category definition, flagging all statements that did not fit the category. Over 400 codes were assigned by the first author, and the second author agreed with 84\% of those code assignments. All discrepancies were reconciled through discussion between the two coders. For example, one common type of discrepancy was related to students' statements about their enjoyment of the project as a whole versus changes in their enjoyment as a result of a particular event. Originally, both types of statements were coded as \emph{affective response}. However, upon discussion, we decided to code the former as instances of students expressing interest in the project itself (\emph{interest and value}), whereas the latter remained coded as \emph{affective response}. {Thus, according to our final scheme, the statement, ``[the project] is more like engineering rather than physics, that's probably the reason why I enjoy it," was coded as an instance of a student expressing interest in the engineering nature of the project. On the other hand, the statement, ``I have a much higher level of enjoyment now that I have solved the stepper motor problem," was coded as an instance of a students' affective response to an event.}

Finally, the first and second authors collaboratively identified emergent subthemes for each \emph{a priori} code category. All subthemes were discussed among the research team as a whole. We describe these and other findings in the following section.


\section{Results}\label{sec:results}

We first report chronological case descriptions for groups A, B, and C, followed by the results of our cross-case analysis. We assign the following pseudonyms to the students in our study: Alan and Avery (group A); Ben, Blake, and Brian (group B); and Carter and Colby (group C). All seven students were men, as were both instructors; hence, we use ``he, him, his" pronouns when referring to students and instructors. {When summarizing students' experiences on their projects, we use minimal technical optics jargon. While some jargon is inevitable, familiarity with optics equipment or techniques is not necessary to understand the case descriptions.}

\subsection{Chronological case descriptions}\label{sec:within-case}
{All three case chronologies have a similar six-part structure: we describe (i) students' background and preparation, (ii)  project goals, (iii) students' initial interest in the project, (iv) two key issues during the project, (v) students' final interest in the project, and (vi) a summary of the case.} In terms of background and preparation, all students had completed 4 to 5 lab courses prior to the start of the project, and most of these lab courses had final projects of their own.


\subsubsection{Group A: Frequency Comb Project}\label{sec:caseA}

Alan and Avery were friends who both had high levels of preparation for the frequency comb project. Both were physics majors who wanted to attend graduate school and engage in experimental atomic, molecular, or optical physics research. Each student had previously engaged in multiple undergraduate research experiences, including one project on which they collaborated with one another. In addition, Alan and Avery had worked together on the frequency comb project during Optics/Lasers a year prior to the start of our study, and Alan had been working on the project during the interim to fulfill his senior thesis requirement.

Group A's project was part of a multi-year effort to develop an ultrafast fiber laser to be used in a high-precision frequency comb. {A frequency comb is laser source that contains multiple, discrete frequencies of light.} Once built, the frequency comb would be used in educational and research contexts in the physics department. To this end, the project had three goals: (i) improve the existing resolution of measurements of the repetition rate of the fiber laser, (ii) broaden the frequency spectrum of the laser pulses, and (iii) measure and stabilize the offset frequency of the frequency comb. Alan and Avery made progress toward the first two goals, but they neither measured nor stabilized the offset frequency of the comb.

When Alan initially began working on the frequency comb the year prior to our study, he wasn't interested in the project:
\QUOTE{Well, when I first started this project last spring in optics I wasn't all that excited about it. \ldots\ I couldn't really decide which project I wanted to do. I turned in my thing late, and [my professor] asked if I would be willing to do this project. I tried this project, and then it turned out I really enjoyed it.}{Alan}{pre-project interview}
During the pre-project interview, after having worked on the project for a year, Alan was able to articulate his interest along multiple dimensions: he was excited to work with Avery and go into ``our own territory," find ``our own solutions," and ``think outside the box." Alan also enjoyed the project because it was a ``cool blend" of conceptual, mathematical, and hands-on physics. Avery's initial interest, like Alan's, was informed by his previous work on the project:
\QUOTE{Since I've been working on the project for such time, it'd be really fulfilling to get it much closer to being done. I feel like we're pretty close.}{Avery}{pre-project interview}

We divide Group A's project into two sequential halves, each lasting about three or four weeks. Each half of the project was defined by a distinct key issue: establishing a mode-lock on the fiber laser (first half), and broadening the spectrum of the pulses (second half).

When establishing a mode-lock, Alan and Avery worked separately on some tasks and together on others. For example, initially, Alan took on the responsibility of realigning optical components while Avery built a circuit to measure the repetition rate of the laser. However, when Avery's circuit wasn't working as expected, Alan helped troubleshoot the circuit. Similarly, once the circuit was functional, Avery helped Alan with the optical alignment. Throughout this process, both students indicated that they felt they could achieve a mode-lock without help from their professor. Nevertheless, the professor provided the students with circuit components, corresponding datasheets, and input on the design of their circuit.

Both students described the process of finding a mode-lock as a frustrating impediment to progress. For example, Alan referred to the process as tedious, unenjoyable, time-consuming, and frustrating:
\QUOTE{Searching for mode-locks using [our method] can be extremely tedious. It is even less enjoyable when it takes many hours of searching when those hours could have been spent making progress on our initial goals of our project.}{Alan}{survey 3, week 4}
\QUOTE{Right away we had a lot of trouble with our laser mode-locking. \ldots\ We spent a couple of weeks just scrambling just to get the project started. That was really frustrating.}{Alan}{post-project interview}
When the mode-lock was finally achieved in the third week, both students indicated that their enjoyment of the project increased. In particular, Avery coupled the frustrating nature of the process to the level of enjoyment he felt:
\QUOTE{Putting time into something that was not giving many signs of life, that wasn't giving clear directions of where to look next, was frustrating and draining. So that makes this case of finding one an ecstatic occasion \ldots\ We can move forward again.}{Avery}{survey 3, week 4}
Alan said his confidence did not change after finding the mode-lock (``We always knew we would [mode-lock the laser]; it just became annoying \ldots"). In contrast, Avery said he felt more confident about the circuit he built than about his ability to achieve a mode-lock:
\QUOTE{It is kind of reassuring to have a component that I suggested, a component that we spent time on to troubleshoot (the photodiode circuit), pay off in the end, but at the same time the mode-lock process we use is so opaque experimentally that if I hadn't happened to twist the dials as they are now, we may have still been there. I am more confident in our tools used to find the beast, but not in my ability to know where to look for it.}{Avery}{survey 3, week 4}

In the second half of the project, Alan and Avery used a method called ``fiber dispersion compensation" to broaden the frequency spectrum of the laser pulses. In doing so, they worked together to clean, polish, splice, and connect optical fibers. Their professor provided them with guidance about which experimental technique to use, provided them with relevant equipment, and taught them relevant skills (e.g., how to connect fibers). At the end of week 5, after trying unsuccessfully for multiple weeks to broaden the pulse spectrum, Alan said he felt ``disappointed" and that the prospects for completing the project seemed ``bleak." One week later, during the final week of the project, Alan and Avery successfully broadened the pulse spectrum. Both students indicated that their enjoyment of the project increased as a result. Each of them viewed this achievement as the first significant step toward completing their project.

During post-project interviews, Alan indicated that his personal interest in the comb had not changed over the course of the seven-week final project; he enjoyed the project from start to finish. He said he enjoyed the project because it gave him a ``sense of discovery" that comes with ``explor[ing] the unknown."  On the other hand, Avery's personal interest did increase, but over the timescale of a year. He said that, when he began working on the project, ``I didn't really know what I was getting into. It was frustrating. Progress in it wasn't coming very quickly." However, the opportunity to continue working on the comb with his friend, Alan, was fun.

When asked whether he had gained insight into why the project may be of interest to others, Alan indicated that, after completing the project, he was able to understand how different aspects of the comb fit together to make a usable tool:
\QUOTE{Rather than just having a couple of cool side projects, it was like an actual---like, we're building a tool, and I'm finally actually kind of seeing it. It was fun to be able to see how this is actually a usable tool that is right on the cusp of being there.}{Alan}{post-project interview}
Similarly, Avery also reported a change in his perception of the project's value to others. As part of his graduate school application process, Avery visited frequency comb facilities at other universities and national labs. These visits helped him situate his project in the context of ongoing research:
\QUOTE{They were talking about how they're using the frequency combs with the atomic clocks. And the incredible, fantastic, precision that they've gotten with those also leveled what I was doing here.}{Avery}{post-project interview}

{In summary, although one member of group A was not initially interested in the frequency comb project, the students expressed interest in the discovery-based and exploratory nature of the project, as well as its utility to other scientists. When establishing a mode lock, each student took control of a different aspect of the project. When broadening the pulse spectrum, the students worked together more closely. Both phases involved periods of frustration, tedium, or disappointment, followed by successful moments that resulted in increased enjoyment of the project. Throughout, the professor supported group A by providing them with relevant equipment and advice. Although the students did not achieve all three project goals, they nevertheless had an overall positive experience.}


\subsubsection{Group B: Scanning Spectrometer Project}\label{sec:caseB}

Ben, Blake, and Brian all described themselves as people who liked solving engineering and applied physics problems. Blake referred to Ben and Brian as ``my buddies." Ben and Blake were both majoring in applied physics, and Brian was majoring in physics and an engineering discipline. All three were interested in pursuing careers in an engineering field. Each student had completed either a summer research experience or an internship, and none of them had previously taken either Optics or Lasers.

Group B's project involved the development a low-cost scanning spectrometer with sufficiently high accuracy and resolution that it could be used to establish a mode-lock for some of the laser systems in Optics/Lasers. To this end, the project had two goals: (i) build a stepper motor system to rotate (or ``scan") a diffraction grating, thus enabling the spectrometer to measure light intensity at different wavelengths, and (ii) calibrate the spectrometer for use with visible and infrared light sources. Ben, Blake, and Brian accomplished the first goal and part of the second; they calibrated and tested the spectrometer in the visible spectrum using mercury and sodium lamps.

At the start of the project, all three students expressed interest in working with microcontrollers, the 3D printer, and other equipment or software. They were excited to work on a hands-on, engineering-type project. For example, Blake sad,
\QUOTE{[The project is] interesting because it uses a lot of different things, I guess, like LabVIEW. We'll use a 3D printer to build some things. It'll require a lot of creativity, I guess.  \ldots\ It is more like engineering rather than physics. That's probably the reason why I enjoy it, or think I'll enjoy it, is because I am more hands-on and more application-based rather than the sheer learning.}{Blake}{pre-project interview}
In addition, Ben said he was excited about the opportunity to work closely with a group of students and his professor:
\QUOTE{I guess it kind of allows us to have some sort of independent---I don't really know how to say what I'm thinking. \ldots\ Kind of smaller group cooperation with the professor, and to get a deeper understanding of how to apply what we've learned in the class.}{Ben}{pre-project interview}

We divide Group B's project into two sequential parts, the first lasting about six weeks and the second lasting about one week. Each part of the project was defined by a distinct key issue: building the stepper motor system (first part), and calibrating the spectrometer (second part).

When building the stepper motor system, Ben, Blake, and Brian divided labor in the following way: Ben designed and built a 3D-printed mount to attach a diffraction grating to the motor; Blake supported Ben by helping troubleshoot problems with the printer; Brian designed and built an electric circuit and corresponding computer code to control the motor via a microcontroller; and all three students collaboratively aligned the relevant optical components and brainstormed solutions to problems. During this phase of the project, Brian expressed concern that ``not all of the group members have put in the same amount of time on the project, which has put more responsibility on myself." Meanwhile, Ben and Blake indicated that the project was a team effort. The professor supported the students by confirming that their alignment was good, answering their questions about the stepper motor system, and ordering parts.

While building the grating mount and circuitry, the students reported feeling ``lost," ``confused," and ``overwhelmed" by the task at hand. However, they all indicated that they thought they could build the system on their own. Once the grating mount and circuitry were built, a functional system was assembled and each student indicated that their enjoyment and confidence increased as a result. For example, Ben said that his enjoyment increased because his stress decreased:
\QUOTE{Before completing the mount, I felt as though we had not made any progress on the project which was quite stressful. Now that we have made progress towards the completion of the project, I feel better able to enjoy the remaining steps in the progress. It's also great to see a clear impact of the work I have done on the project's completion.}{Ben}{survey 3, week 4}
Although the system was functional, the step-size of the motor rotation was too coarse for use in the spectrometer. Increasing the rotational resolution of the motor was a harder task than the students initially anticipated. As the end of the semester approached, they reported feeling frustrated, unsure, and ``like no more progress can be made." For example, Blake wrote,
\QUOTE{I was a little frustrated, because getting this part of the project working determines everything else we did.}{Blake}{survey 2, week 6}

In the last week of the semester, after soliciting ideas and advice from their professor, the students successfully decreased the step size of the motor. They then assembled, calibrated, and tested the spectrometer. Both Ben and Brian indicated that this achievement significantly improved their enjoyment of the project and their confidence (Blake did not complete survey 3 in week 7). For example, Ben referred to the calibration and testing of the spectrometer as a ``huge boost in morale":
\QUOTE{Reaching that goal was a large boost in morale. It also helped us see how close we are to our final project goals. We all felt more confident after these solutions were found as we all played a large part in coming up with the overall answer to a significant issue we were facing.}{Ben}{survey 3, week 7}
In his post-project interview, Blake said that collecting data with the spectrometer was memorable because he ``spent a lot of time aligning it and getting it focused and where we could actually see data." Brian noted that overcoming challenges on the spectrometer project improved his confidence in engineering and physics more generally:
\QUOTE{The spectrometer project was a challenge different than most lab projects I have worked on. \ldots\ There were more problems that occurred in this project than most I had encountered in the past, and figuring out how to get past those was often difficult. Overall, I now feel as though I have a lot more confidence in my ability to get past difficult engineering/physics problems on my own.}{Brian}{survey 3, week 7}

During post-project interviews, both Ben and Brian indicated that their interest in the project increased over time. Both students attributed their increased interest to a better understanding of the purpose of the project. For example, Brian said,
\QUOTE{So now, especially after I've completed it, I understand its value, seeing the different prices of spectrometers and understanding how a spectrometer works and having to be so flexible for adjusting it to different wavelengths. So I understand why we did the project better now compared to at the beginning.}{Brian}{post-project interview}
Blake said his interest did not change; he enjoyed the engineering nature of the project from start to finish. However, like Brian, Blake also had a new appreciation of the relevance of their project. In particular, Blake noted that the spectrometer could be useful for other projects in Optics/Lasers:
\QUOTE{We got told by a few of the other groups that it would be nice if they could use it once we were done. \ldots\ I definitely can see how it's useful.}{Blake}{post-project interview}

{In summary, all members of group B were initially interested in the spectrometer project because it was an opportunity to solve an engineering-style problem using a hands-on approach and sophisticated equipment. Over the course of the project, all group members developed an increased appreciation for the spectrometer's utility in the Optics/Lasers course. The bulk of their project was spent building a stepper motor system, a long process during which students each took control of a different aspect of the project. While struggling to build the system, the students felt frustrated and overwhelmed. When the stepper motor was finally built and the spectrometer was calibrated, two group members reported large increases in their enjoyment of the project and their confidence with solving engineering/physics problems. Throughout, the professor supported group B by providing them with equipment, advice, and validation. Although the students did not calibrate the spectrometer for use with infrared sources, they nevertheless had an overall positive experience.}


\subsubsection{Group C: Plasmon Laser Project}\label{sec:caseC}

Carter and Colby were majoring in physics and applied physics, respectively.  Neither had previously participated in undergraduate research. Carter was considering the possibility of applying to graduate school; Colby was considering a career in technical sales. However, both students expressed a lack of clarity about their own interests and passions with respect to future education and career pathways.

Group C's project involved the development of a surface plasmon laser, or spaser. A spaser is similar to a laser, but with plasmons playing the role of photons and a nanoparticle playing the role of the laser cavity. The project had three goals: (i) fabricate the gain material for the spaser, (ii) characterize the spectra of different gain media, and (iii) achieve lasing. Carter and Colby accomplished the first two goals, but not the third.

At the start of the project, both students said they were excited to work on something ``novel," and both expressed concern about their lack of relevant content knowledge. Carter said he was ``not normally a hands-on kind of person," but he was looking forward to getting hands-on experience with the project. In particular, he was excited about his access to the nanotechnology facilities in his department:
\QUOTE{I think it sounds like something that's really new, really novel, and I think that that's really cool. I like [the professor] as a professor a lot. I would like to work in the nano lab.}{Carter}{pre-project interview}
Colby was not initially interested in the spaser project. Nevertheless, he expressed interest in learning more about spasers and feeling ownership of the project:
\QUOTE{To be honest, I think it'll take a little bit of learning to find some interest in it. I mean we ranked our choices and this one didn't totally stand out. I like the idea of trying to make something else lase. I think that's kind of an interesting concept that's a little bit foreign to me. \ldots\ I guess [I'm] a little bit excited about learning the material and hopefully feeling ownership of something.}{Colby}{pre-project interview}

We divide Group C's project into two sequential halves, each lasting about three or four weeks. During the first half of the project, Carter and Colby aligned the lasers that would later be used to characterize the spectrum of their samples. During the second half, Carter and Colby each worked on distinct aspects of the project. Colby fabricated different types of gain media, and Carter characterized the spectra of those media. In their post-project interviews, both students said that they enjoyed this division of labor. For example, Colby said,
\QUOTE{One piece that I enjoyed was I kind of took---me and my lab partner each had a piece that we took a little bit of control over. For me, I worked in the clean room doing some chemistry in the fabrication of the gain material and the spasing material. So for me that was kind of what I felt was a little bit memorable, was actually creating the chips and feeling pretty productive in that sense.}{Colby}{post-project interview}

When aligning the lasers, Carter and Colby worked together as a pair. Their professor provided them with information about how to use the equipment, explained the alignment process, and confirmed that the they set up their optical components correctly. Both students indicated that they felt they could align the lasers on their own, without further help from the professor. Colby described this process as tedious, whereas Carter said that he enjoyed the ``miscellaneous setup" because he liked working toward a goal.

Once the optical setup had been aligned, Colby began working in the nanotechnology lab to fabricate different gain media. The professor gave Colby a tutorial on the fabrication process, and also provided guidance about which ingredients to try when fabricating media. Colby tried three different recipes that used various combinations of dye powders, solvents, and coagulants. Meanwhile, Carter was performing spectral analyses on Colby's samples, using both incoherent and coherent light sources. One challenge that Carter encountered was that fluctuations in the irradiance of the samples made it difficult to characterize their spectra. The students decided to measure the irradiance using a power meter rather than a spectrometer; the power meter output was more stable since it averaged over fast fluctuations in light intensity. Carter said that the decision to use the power meter ``was our idea;" the professor simply confirmed that the decision was appropriate.

Ultimately, though they did not observe lasing, Carter and Colby were able to determine that the nanoparticles on one of their samples was behaving like a cavity. Colby indicated that his enjoyment and confidence did not change as a result of this accomplishment:
\QUOTE{[My level of enjoyment] didn't change significantly because it was one of the final measurements of our lab and we did not see the desired result that we set out to achieve initially. \ldots\ [My level of confidence] was relatively unchanged because although it was a decent result it wasn't what we desired so I did not feel like we solved all the problems that we would have liked to have solved.}{Colby}{survey 3, week 7}
Carter, on the other hand, said that ``finally getting results \ldots\ was exciting." In his post-project interview, he elaborated on his feelings:
\QUOTE{I think the biggest thing that I remember is when we were trying to characterize our gain material. We probably spent a week, week and a half, maybe two weeks, trying to get it to work with a spectrometer, and eventually we had to just say this isn't working and we did it with a power meter instead. So there was a lot of buildup, and then when it worked with the power meter that was a really memorable moment.}{Carter}{post-project interview}
Moreover, Carter said that the major outcome of the project as a whole was his improved confidence:
\QUOTE{I think I'm a lot more confident in my ability to solve problems on my own. I wouldn't, I don't think, have felt comfortable going into an internship or a job in physics because everything's been in the classroom. But having this opportunity to really get into problem solving with help from a professor, but also a lot on my own, was really helpful.}{Carter}{post-project interview}

During post-project interviews, both students indicated that their interest in the project increased from start to finish. Going beyond the context of the project, Carter indicated that his interest in optical physics had increased:
\QUOTE{I think I would never before this class have thought that I would want to work with lasers, but a lot of the stuff that I've gotten to work with through the project has been really cool, so I think I'm more interested in this kind of field than before.}{Carter}{post-project interview}
In contrast, Colby described only a slight increase in his interest in the project:
\QUOTE{I would say slightly. I think originally when I got the project I had no idea what it was. I don't know that I found a passion for the particular project, but I did enjoy learning more about a particular subject. I don't know that I have a thirst for going deeper into the surface plasmon lasing---that's what we worked on---but I did gain a thirst for learning a little bit more about it. I thought that was kind of fun.}{Colby}{post-project interview}

{In summary, although one member of group C was not initially interested in the surface plasmon laser project, both students expressed excitement to learn about a new area of physics research. By the end of the project, both students reported an increase in their interest in the project and related physics. The students worked together when aligning lasers in their optical setup, but they worked separately to fabricate and characterize different gain media during the second half of the project. Throughout, the professor supported group C by providing them with tutorials, advice, and validation. The students did not achieve the ultimate goal of their project (i.e., lasing), which resulted in mixed affective responses among the pair: one student said that his confidence and enjoyment did not increase, whereas the other experienced increases in both of these dimensions. Both students had an overall positive experience on the project.}


\subsection{Results from cross-case analysis}\label{sec:cross-case}

For each student, and hence each group, we identified multiple instances of each of the five dimensions of ownership in our \emph{a priori} coding scheme (Sec.~\ref{sec:cross-case-methods}): student agency, instructor mentorship, peer collaboration, interest and value, and affective response. In this section, we organize the results of our cross-case analysis according to three emergent themes: \emph{division of labor and collaborative brainstorming}, \emph{development of interest over time}, and \emph{cycles of struggle and success}. The first theme is related to the dimensions of agency, mentorship, and collaboration. The second theme is related to interest and value, and the third theme is related to students' affective responses.

\subsubsection{Division of labor and collaborative brainstorming}

The first cross-case theme that emerged from our data was that students struck a balance between agency, mentorship, and collaboration by dividing labor among themselves while working together with each other and the professor to brainstorm solutions to problems.

The most common forms of student agency involved student participation in creating plans and strategies, as well as setting project goals. The former accounted for about half of all instances of student agency in our data set, the latter accounted for about a quarter. Other instances of student agency included times when students were reflecting on their progress or managing their time and effort on the project. Students rarely described instances when they were the only person responsible for creating a plan or setting a goal. Often, plans and goals were also informed by other students or input from their professor.

We identified four types of instructor mentorship: setting large-scale project goals, confirming that a result is correct, providing students with equipment and corresponding background information, and working with students to troubleshoot problems that inevitably arise. For example, the professor provided group B with both strategies and equipment to help them overcome problems with their stepper motor:
\QUOTE{Our professor helped my group think through different solutions to decreasing the step size of our stepper motor. He helped formulate some ideas for our circuit, and also put an order in for some electrical components that could help solve the problem.}{Brian}{survey 2, week 6}

We also identified four types of student collaboration: collaboratively working on the same task, collaboratively learning relevant background information, collaboratively brainstorming solutions to problems, and dividing labor among group members. Dividing labor was often described in combination with collaborative brainstorming. For example, Alan said,
\QUOTE{Many of the tests on the circuit really only could have person testing them. Therefore, [Avery] did a lot of the troubleshooting of the circuit while I continued our search for a mode-lock using the much slower [optical spectrum analyzer]. However, when it came time to brainstorm and make changes to the fibers, we both did significant work.}{Alan}{survey 2, week 3}
Similarly, Ben, Blake, and their professor helped Brian diagnose problems with the circuit he was building for the stepper motor system. And, despite dividing the tasks of fabricating and characterizing samples, Carter and Colby worked together to solve the problem of the samples' fluctuating irradiance.

Dividing labor gave each student a chance to have control over one particular aspect of the project. At the same time, collaborative brainstorming allowed for students to be invested in all aspects of the project, and for the professor to support students' progress during times when they might otherwise feel stuck.

\subsubsection{Development of interest over time}

The second cross-case theme that emerged from our data was that students' interest in the project was cultivated over time, even in cases where students were not initially excited about the project. {Our analysis showed that students found their projects interesting or valuable because their projects were discovery-based, involved new physics, included engineering-style challenges (e.g., making an apparatus cheaper or smaller), or provided students with access to equipment and facilities that they wanted to use. However, these patterns varied by group and over time.}

Most, but not all, students were initially interested in their projects. Alan and Colby both described being assigned to projects that were not their first choices. (For Alan, the initial project assignment happened a year prior to the start of our study.) Nevertheless, both students developed interest over time. Indeed, almost all students described an increase in their interest in the project from start to finish. The only exception was Blake, whose interest in the spectrometer project remained high throughout.

The reasons for students' initial interest in the project varied from group to group. The students in group A were interested in the frequency comb project in part because they had been working on it for a long time, and they wanted to see it through to completion. The students in group B were interested in their project in part because it was ``engineering-based," and because it gave them the opportunity to work with 3D printers and microcontrollers. In group C, Carter was interested in working on novel physics. Like the students in group B, he was also excited about gaining access to particular technology---in his case, the nanotechnology lab.

In groups A and B, increased interest in the project was coupled to a better understanding of its value to others. These students said that their interest in the project increased as they gained a better understanding of the purpose, value, and utility of their projects. The students in group C described a different mechanism for their increased interest in the project: they developed desire to learn more about lasers and optics generally (Carter) and spasers more specifically (Colby).

\subsubsection{Cycles of struggle and success}

The third and last cross-case theme that emerged from our data was that students experienced multiple cycles of extended struggle that culminated in momentary successes. These successful moments contributed to students' increased enjoyment of the project, and sometimes also to increases in their confidence with respect to their ability to overcome project-related challenges.

{Our analysis of students' affective responses revealed multiple patterns. For example, students described feeling confident more often than they described a lack of confidence. Sometimes, students' confidence (or lack thereof) was connected to their familiarity with a particular aspect of the apparatus. For example, Ben and Carter coupled their confidence on certain design and analysis tasks to their familiarity with the relevant software, whereas Brian expressed concern about his lack of familiarity with stepper motor operation. On the other hand, students described feelings of enjoyment about as frequently as they described feelings of frustration and stress. For each student, their confidence, enjoyment, and frustration varied dynamically as they cycled between periods of struggle and moments of success.}

Periods of struggle were characterized by feelings of frustration, tedium, overwhelmedness, or a general sense that no progress was being made on the project. When describing these periods, students often referred to the temporal duration of their struggle. For example, when Alan described the process of achieving a mode-lock as frustrating and tedious, he also noted that it required ``many hours of searching" and ``a couple of weeks [of] just scrambling." Similarly, in week 6, when group C was trying to overcome the challenge of irradiance fluctuations, Carter said he felt that his group's efforts ``seemed to have been for nothing." During his post-project interview, Carter recalled that he and Colby ``spent a week, week and a half, maybe two weeks" trying to resolve the issue.

In all three groups, these extended periods of struggle culminated in a successful moment during which a challenge was overcome. Successful moments were characterized by feelings of excitement, accomplishment, or a general sense that progress could finally be made on the project. Students also characterized these moments as ``nice," ``cool," ``good," or ``great." For some students, it was important to see the tangible impacts of their own contributions to the project. For example, when group A finally achieved a mode-lock, Avery described the moment as an ecstatic occasion because his group could ``move forward again," and he specifically identified the circuit he built as playing a role in the successful moment. Similarly, when group B finally assembled the stepper motor system, Ben said he was ``better able to enjoy" the project because his group was making progress on the project. Like Avery, Ben also noted that his specific contribution (the 3D-printed grating mount) played a role in achieving a particular goal.

In many cases, successful moments were accompanied by an increase in students' confidence about their own, or their group's, ability to overcome challenges on the project. However, not all successful moments resulted in an increase in confidence. For example, after group A achieved a mode-lock, Alan said his confidence was unchanged because he always knew that he and Avery could accomplish this task: ``We have found mode-locks in the past, but sometimes it takes more searching than others." Similarly, when group C finally observed evidence that their gain medium was behaving like a spaser cavity, Carter said,
\QUOTE{The alignment and analysis that I did for my part of this successful moment was fairly similar to what I have done earlier in the project. so while I am proud of myself for having done it, it isn't as edifying to solve the same problem a second or third time.}{Carter}{survey 3, week 7}
For both Alan and Carter, the task they accomplished was challenging and time-consuming, but it was also something they had done in the past. Therefore, while the accomplishment resulted in increased enjoyment of the project, it did not improve their (already high) confidence in their ability to solve the problem. Alan's and Carter's reactions are consistent with the findings of Bandura~\cite{Bandura1997} regarding repeated mastery experiences and changes in self-efficacy.

Finally, we note that none of the groups achieved all of the initial goals of the project: group A did not stabilize the offset frequency of the comb, group B did not calibrate their spectrometer in the infrared spectrum, and group C did not achieve lasing. Nevertheless, all groups achieved some of their project goals, experiencing multiple cycles of extended struggle and momentary success along the way. This suggests that failure to meet an overarching project goal may not negatively impact students' sense of ownership, provided they experience struggle and success on one or two key milestones.


\section{Discussion}\label{sec:discussion}

Because each participant in our study described all five dimensions of ownership, we argue that the students in all three groups felt that they had ownership of their projects. In addition, our analyses yielded three themes about student ownership of final projects in an optics lab course: (i) coupling division of labor with collective brainstorming can help balance student agency, instructor mentorship, and peer collaboration; (ii) initial student interest in the project topic is not always a necessary condition for student ownership of the project; and (iii) student ownership is characterized by a wide range of emotions that fluctuate in time as students alternate between extended periods of struggle and moments of success while working on their project. These themes constitute the major findings of this work. {In this section, we discuss limitations of our findings, implications for instruction, and ideas for future research.}

\subsection{Limitations}

{We highlight two limitations of the study. First, our participant pool was relatively homogenous with respect to race, ethnicity, and gender: almost all students who complete Optics/Lasers are white men. Moreover, each case we chose for analysis consisted of students with similar educational backgrounds and interests. Therefore, our study does not provide insight into the complex dynamics that likely arise among more heterogeneous student groups. For example, Grover et al.~\cite{Grover2017} recently showed that, for groups of students collaboratively solving math problems, the social cohesion of a group---i.e., a composite measure of the extent to which a group member feels like they belong and the extent to which one group member enjoys another's presence---depends on the group's gender composition. Because the details of social interactions among students and their professors are an important aspect of student ownership, our study likely has limited generalizability to more heterogeneous learning environments. }

{Second, we limited our analysis to groups for which all students had similarly positive experiences during the final project. As a result, our study does not provide deep insight into the distinction between individual and group ownership. This distinction has been studied in other physics contexts. For example, Enghag and colleagues~\cite{Enghag2006,Enghag2008,Enghag2009} found that a group may collectively have ownership of a project even if some individual group members do not. Therefore, our findings may have limited generalizability to groups in which students have varied engagement and outcomes.}

\subsection{Implications for instruction}

Our analyses suggest multiple teaching principles that can inform the design and framing of physics lab projects for which student ownership is a desired learning outcome. These teaching principles reinforce recommendations made in other educational contexts; we note synergies as appropriate.

Projects with multiple subtasks that can be performed in parallel could promote student ownership by facilitating division of labor and collaborative brainstorming among students. The findings of Rainer and Matthews~\cite{Rainer2002} suggest that dividing labor in a way that leverages each student's particular expertise is an effective strategy. However, students should not be isolated; encouraging group members to work together to troubleshoot problems with each other's subsystems can help ensure that each student feels connection to the project as a whole. Moreover, instructor participation in the process of brainstorming solutions to problems could be a way for instructors to provide the authoritative-but-restrained guidance described by Mikalayeva~\cite{Mikalayeva2016} and Hanauer and colleagues~\cite{Hanauer2012,Hanauer2014}---especially if the responsibility for choosing and enacting a particular solution strategy remains with the students~\cite{Savery1996}.

Although Milner-Bolotin~\cite{Milner-Bolotin2001} showed that students' initial interest was correlated with a sense of ownership at the start of the project, our results suggest that a lack of initial interest is not necessary for students to develop ownership as the project progresses. Coupled with Milner-Bolotin's finding that students' level of choice of project topic was not correlated with ownership at the end of the project, one might erroneously conclude that the project topic plays a negligible role in students' development of a sense of ownership. However, in our study, students' interest changed over time due, in part, to their improved understanding of the purpose of the project. Similarly, others have argued that ownership is cultivated when students work on projects that are purposeful~\cite{Rainer2002} or socially/professionally relevant~\cite{Hanauer2012,Hanauer2014}. Therefore, we argue that projects that are useful or relevant to others can foster a sense of student ownership. Moreover, instructors can support the development of student ownership over time by helping students see the importance of their work as they become more familiar with the details of the project.

Finally, projects that are designed to minimize student frustration (e.g., by minimizing the need to troubleshoot equipment or iterate on designs) may be inappropriate for supporting student ownership. Instead, a project that has ambitious-but-achievable goals may be more ideal. Our results suggest that it is not necessary for students to accomplish the overarching project goal, provided that they make tangible progress on one or more challenging subgoals. This finding is consistent with the work of Hanauer and colleagues~\cite{Hanauer2012,Hanauer2014}, who noted that experiences of challenge and satisfaction play an important role in students' development of a sense of project ownership. In addition, like Milner-Bolotin~\cite{Milner-Bolotin2001}, we also observed dynamic fluctuations in students' affective responses to their projects. Beyond articulating multiple project subgoals, instructors could also frame frustration, tedium, worry, uncertainty, relief, excitement, and other affective responses as normal, necessary, and interrelated aspects of doing physics---and hence of completing the project (cf. Ref.~\cite{Eblen-Zayas2016}). Doing so could help students anticipate and regulate their own emotional responses to setbacks and successes, which are important components of agency~\cite{Bandura2008}. Moreover, framing the project this way could help students become familiar with the emotional realities of doing science, which, as Jaber and Hammer~\cite{Jaber2016a,Jaber2016b} have argued, is critical to students' development of scientific expertise.

\subsection{Ideas for future research}

{
Other work has drawn connections between ownership and positive affect, like excitement or pride~\cite{Wiley2009,Hanauer2014,Little2015,O'Neill2005}. While some studies acknowledge that students experience frustration during their projects, frustration is usually framed as something that needs to be overcome~\cite{Hanauer2012,Hanauer2014} or that explains low levels of ownership~\cite{Milner-Bolotin2001}. We have previously argued that students' affective responses while working on projects in Optics/Lasers are complex and dynamic, and that a balance of both positive and negative responses might support, rather than hinder, students' development of a sense of ownership~\cite{Stanley2016}. In this study, we have shown not only how and why students' affective responses fluctuate between frustration or disappointment and excitement or relief, but we also have argued that these responses are interdependent: long periods of struggle are sometimes \emph{the reason} that successful moments are joyous occasions. Little~\cite{Little2015} has described similar dynamics in the context of students' sense of pride, suggesting that some amount of struggle can be desirable for many reasons. Nevertheless, it is reasonable to assume that long periods of struggle can be \emph{too} long; eventually, students may lose interest in a project that yields few significant results. Future research could explore in more detail the ways that struggle facilitates, or inhibits, students' sense of project ownership: how much struggle is too much (or too little)?
}

{Another potential avenue for future exploration involves student ownership in groups comprised of students with highly varied levels of interest in, commitment to, or control over a shared project. Contrasting the experiences of different students in such groups could contribute to the development of a more comprehensive conception of student ownership. In particular, this work could build on that of Enghag and Niederrer~\cite{Enghag2008}, who distinguished between group and individual ownership, i.e., a group's level of control over the management and execution of the project, versus the extent to which the group incorporates a particular student's input. This distinction could provide additional insight into the ways that division of labor and collaborative brainstorming facilitate student ownership: how are tasks assigned, and whose ideas are taken up (or not taken up) during brainstorming sessions? Distinguishing between group and individual ownership may also be a useful approach for understanding student ownership in groups with either a domineering or disengaged student. Such groups may be characterized by challenging social interactions among peers and mentors. However, it is unclear whether extended periods of \emph{social} struggle contribute to student ownership in similar ways to periods of \emph{technical} struggle, either at the individual or group level.}

{Exploring these and other questions will not only help develop a more comprehensive understanding of what student ownership is and how it can be supported, but will likely also result in additional practical guidelines for instructors for whom ownership is a learning goal.}


\section{Summary and next steps}\label{sec:conclusion}

{We investigated student ownership of multiweek final projects in an upper-division optics lab course using a multiple case study methodology. Interview data, survey data, and course artifacts were collected for 12 groups of students over the course of 2 semesters. Three student groups were chosen for analysis. We reconstructed chronological descriptions of each group's project. In addition, we analyzed interview and survey data using an \emph{a priori} analysis scheme based on five dimensions of student ownership articulated by Hanauer and colleagues~\cite{Hanauer2014,Hanauer2012}. Three emergent subthemes were identified: (i) coupling division of labor with collective brainstorming can help balance student agency, instructor mentorship, and peer collaboration; (ii) initial student interest in the project is not always a necessary condition for student ownership of the project; and (iii) student ownership is characterized by a wide range of emotions that fluctuate in time as students alternate between extended periods of struggle and moments of success while working on their project.}

{This work extends the literature on student ownership into a new domain, namely, upper-division physics labs. It also complements the existing literature on teaching and learning in optics courses; whereas previous work has focused predominantly on skills and concepts, our work draws attention to affective and social dynamics in the instructional optics lab. Going forward, our findings pave the way for developing effective, research-based practices for the design and implementation of final projects in physics labs. To that end, our planned future work involves investigating the dynamics of student ownership in multiple institutional contexts. This broader investigation will help us characterize additional features of learning environments, instructional strategies, and social interactions that support (or hinder) students' sense of ownership of their final projects.}


\begin{acknowledgements}
{The authors acknowledge Chad Hoyt, Angela Little, Laura Kiepura, and Gina Quan, respectively, for assistance with data collection, study design, interview transcription, and feedback on a draft of this manuscript. This work also benefitted from useful discussions with the PER groups at University of Colorado Boulder and University of Maryland College Park.} This material is based upon work supported by the NSF under Grant Nos. DUE-1323101, DUE-1726045, PHY-1734006, PHY-1560023, and PHY-1208930.
\end{acknowledgements}


\bibliography{./_subfiles/ownership_database_2016}

\end{document}